\documentclass[%
reprint,
showpacs,
 amsmath,amssymb,
 aps,
prb,
floatfix,
]{revtex4-1}

\usepackage{graphicx}
\usepackage{dcolumn}
\usepackage{bm}
\usepackage[colorlinks,linkcolor={blue},citecolor={blue},urlcolor={blue}]{hyperref}
\usepackage{multirow}
\usepackage{array}
\usepackage{booktabs}
\usepackage{ctable}
\usepackage{upgreek}
\usepackage{epsfig}
\usepackage{mathrsfs}
\usepackage{amssymb}
\usepackage{amsbsy}
\usepackage{color}
\usepackage{cancel}
\usepackage{pifont}
\usepackage{marginnote}
\usepackage{dsfont}

\newcommand{\ci}[2][\sigma]{C_{#2#1}}
\newcommand{\cidag}[2][\sigma]{C_{#2#1}^{\dag}}

\newcommand{\eps}{\epsilon}

\newcommand{\up}{\uparrow}
\newcommand{\dn}{\downarrow}

\renewcommand{\vec}[1]{{\boldsymbol{#1}}}
\newcommand{\sg}{\sigma}

\newcommand{\bcen}{\begin{center}}
\newcommand{\ecen}{\end{center}}
\newcommand{\btab}{\begin{tabular}}
\newcommand{\etab}{\end{tabular}}
\newcommand{\bdes}{\begin{description}}
\newcommand{\edes}{\end{description}}

\newcommand{\beq}{\begin{equation}}
\newcommand{\eeq}{\end{equation}}
\newcommand{\bea}{\begin{eqnarray}}
\newcommand{\eea}{\end{eqnarray}}

\newcommand{\bary}{\begin{array}}
\newcommand{\eary}{\end{array}}
\newcommand{\benum}{\begin{enumerate}}
\newcommand{\eenum}{\end{enumerate}}
\newcommand{\bitem}{\begin{itemize}}
\newcommand{\eitem}{\end{itemize}}

\newcommand{\bsig}{\mbox{\boldmath $ \sigma $}}

\newcommand{\btau}{\mbox{\boldmath $ \tau $}}

\newcommand{\bdelta}{{\boldsymbol{\delta}}}
\newcommand{\bTheta}{{\boldsymbol{\Theta}}}

\newcommand{\bk} { \bm{k} }
\newcommand{\br} { \boldsymbol{r}}

\newcommand{\bx} { \mbox{\boldmath $x$}}

\newcommand{\bzero} { \mbox{\boldmath $0$}}

\newcommand{\dou}{\partial}

\newcommand{\D}[1]{\mbox{d}{#1}}

\newcommand{\ket}[1]{| #1 \rangle}

\newcommand{\Itwo}{{\mathds{1}}}
\newcommand{\Hds}{{\mathds{H}}}

\newcommand{\eqn}[1] {eqn.~(\ref{#1})}

\newcommand{\Fig}[1]{Fig.~\ref{#1}}

\newcommand{\myfigwidth}{0.9\columnwidth}

\newcommand{\mylabel}[1]{\label{#1}}

\newcommand{\myonlinecite}[1]{[\onlinecite{#1}]}

\begin{document}


\title{Continuum Theory of Edge States of Topological Insulators: Variational Principle and Boundary Conditions}

\author{Amal Medhi}\email{amedhi@physics.iisc.ernet.in}
\author{Vijay B.~Shenoy}\email{shenoy@physics.iisc.ernet.in}
\affiliation{Center for Condensed Matter Theory, Indian Institute of Science, 
Bangalore 560012, India}



\date{\today}

\begin{abstract}
  We develop a continuum theory to model low energy excitations of a generic four-band time reversal invariant
  electronic system with boundaries. We propose a variational energy functional for the wavefunctions which allows us derive natural boundary conditions valid for such systems. Our formulation is particularly suited to develop a continuum theory of the protected edge/surface excitations of topological insulators  both in two and three dimensions. By a detailed comparison of our analytical formulation with tight binding calculations of ribbons of topological insulators modeled by the Bernevig-Hughes-Zhang (BHZ) hamiltonian, we show that the continuum theory with the natural boundary condition provides an appropriate description of the low energy physics. As a spin-off, we find that in a certain parameter regime, the gap that arises in topological insulator ribbons of finite width due to the hybridization of edges states from opposite edges, depends non-monotonically on the ribbon width and can nearly vanish at certain ``magic widths''. 
\end{abstract}


\maketitle

\section{Introduction}
\mylabel{sec:Intro}

One of the physically observable phenomena in
topological insulators (TI) is the existence of the linearly
dispersing gapless edge (in two dimension (2D) or surface (in three
dimension (3D)) states which are topologically protected against
moderate electronic interactions or nonmagnetic
disorder.\cite{KanePRL05a,KanePRL05b,BernevigSCI06,HasanRMP10,ZhangRMP2011} The
fact that these conducting edge states can host spin current without
dissipation makes topological insulators (TI) a promising
candidate in technological
applications.\cite{MooreNAT10,KonigSCI07} Understanding the
nature of the edge states has been an aspect of interest in theoretical studies of topological
insulators.\cite{KonigJPSJ08,ZhouPRL08,ImuraPRB10,MaoJPSJ10,MaoPRB11}

Properties of edge states can be studied by constructing appropriate tight binding
model Hamiltonians of TIs and examining their eigenstates for lattices
with boundaries. An alternative route is to construct a low
energy continuum theory\cite{BernevigSCI06,ZhouPRL08,HohenadlerPRL11} that 
allows for analytical treatment that aids the development of field
theoretic description in the presence of
interactions.\cite{ZhangPRLPaper,MooreWuPRBPaper} Such approaches have
been gainfully employed earlier in studies of
graphene.\cite{BreyFretig,Akhmerov2008,NetoRMP2009,BhowmickShenoy,vanOstaay2011} In the
analytic calculation, the edge states are obtained by subjecting
appropriate boundary condition (BC) on the wavefunction.  Here one 
usually\cite{KonigJPSJ08,ZhouPRL08,ImuraPRB10,MaoJPSJ10,MaoPRB11}
imposes the {\em fixed} boundary condition (also called as {\em
  essential} or Dirichlet boundary condition in the mathematical
literature\cite{Gelfand1965}) where the wavefunction
is assumed to be zero at the boundaries or at a fictitious layer of
atoms just outside the boundaries. Such a choice of BC constraints the
nature of the wavefunction in that the maximum weight of the edge
state does not occur in the edge layers but in bulk layers that are
near the edge layer. In the presence of interactions, the edge states
and bulk states mix and the ensuing physics is determined crucially by
this mixing. In a recent study\cite{MedhiARX2011}, it was shown
that the Mott transition in topological insulator ribbons can occur in
two different routes -- the synchronous and asynchronous routes --
depending on the nature of edge states. A continuum field theoretic
analysis of such a phenomenon, therefore, requires a careful treatment
of the edge states so that their profile correctly captures the mixing
with the bulk states.

With this motivation, in this paper, we develop a continuum theory of
time reversal invariant four-band model Hamiltonians that have been extensively
used in the analysis of topological insulators in two and three
dimensions. We construct an energy functional of the wave functions;
the wave function that renders this energy functional extremum is
shown the satisfy a stationary Schr\"odinger equation that matches the
four-band lattice theory at long wavelengths. As a key outcome of this
approach, we derive a new boundary condition, the {\em natural}
boundary condition.\cite{Gelfand1965} This boundary condition is valid for
any four-band time reversal invariant system in two and three
dimensions. We use the BHZ model\cite{BernevigSCI06} that has been studied
earlier\cite{KonigJPSJ08,ZhouPRL08,ImuraPRB10,MaoJPSJ10} to show that wthin a regime of parameters of this
model, the natural boundary condition provides an excellent
description of the edge states. In the process of this study, we show
that the gap that arises from the hybridization of the edge states
localized on the opposite edges of a ribbon is a non-monotonic function
of the ribbon width. This finding could potentially be useful in
many applications such as design of thermoelectric devices
etc.\cite{MurakamiPRB2010,MoorePRL2010}

In the following section (sec.~\ref{sec:FourBand}) we introduce a
general four-band lattice Hamiltonian that is time reversal
invariant. Sec.\ref{sec:ContinuumTheory} contains the continuum theory
of these systems, the formulation of a variational principle and
derivation of the boundary conditions. A detailed comparison of the
numerical tight binding calculations and the analytical continuum
theory is carried out in sec.~\ref{sec:BHZ} using the BHZ
model,\cite{BernevigSCI06} in its topological regime. The paper is
concluded in sec.~\ref{sec:Discussion} which contains a discussion,
significance and summary of the results.

\section{Four-Band Time Reversal Invariant Systems}
\mylabel{sec:FourBand}

Consider a Bravais lattice in two or three dimensions whose sites are
labelled by $I$. Each lattice site has two orbitals (or ``basis''
sites such as A-B sites in the graphene lattice, sometimes also referred to as ``flavours'') labelled by
$\alpha$. The operator $C^\dagger_{I \alpha \sigma}$ creates an
electron of spin $\sigma$ (quantized along some convenient axis) in
the orbital $\alpha$ at site $I$. The Hamiltonian of the system is
given by
\beq\mylabel{eqn:TBHamiltonian}
{\cal H} = -\sum_{I \bdelta} t_{\alpha\sigma, \beta\sigma'}(\bdelta) C^\dagger_{(I + \bdelta) \alpha \sigma} C_{I \beta \sigma'}
\eeq
where $\bdelta$ runs over lattice vectors, summation over repeated
orbital and spin indices is implied.  The hopping matrix elements
$t_{\alpha \sigma, \beta\sigma'}(\bdelta)$ are such that the
Hamiltonian \eqn{eqn:TBHamiltonian} is time reversal invariant. Hamiltonians discussed in the literature on topological
insulators\cite{HasanRMP10,ZhangRMP2011} are of this type.

With the aim of developing a long wavelength continuum theory of such
systems, we cast the Hamiltonian in the reciprocal space:
\beq\mylabel{eqn:TBkspace}
{\cal H} = \sum_{\bk \in {\cal B}} H_{ab}(\bk) C^\dagger_{\bk a} C_{\bk b}
\eeq
where $a$ (and $b$) is an index that represents the composite $\alpha
\sigma$. Repeated $a$ and $b$ indices are summed over and $\bk$ runs
over ${\cal B}$, the Brillouin zone which is a torus for 2D systems
and a 3-torus in 3D systems. Following
Refs.~\myonlinecite{MurakamiSCI2003,MurakamiPRB04,KanePRL05a}, we now write the
matrix $H(\bk)$ in a basis of sixteen $4\times4$ matrices, broken up
into two groups $\Gamma^m$ $(m=0-5)$ and $\Lambda^n$ $(n=1-10)$, i.e,

\beq \mylabel{eqn:Hexpand}
H(\bk) = \sum_{n=0}^5 d_n(\bk) \Gamma^n + \sum_{m=1}^{10} e_m(\bk) \Lambda^m
\eeq
where $d_n(\bk)$ and $e_n(\bk)$ are smooth functions of $\bk$. The
matrices $\Gamma$ and $\Lambda$ are defined using $\btau$ and $\bsig$,
the $2\times 2$ Pauli matrices associated with the orbital and spin
degrees of freedom, and $\Itwo$, the $2\times2$ identity matrix. We
have, $\Gamma^0 = \Itwo \otimes \Itwo$,

\beq\mylabel{eqn:GamDef}
\Gamma^{1,2,3,4,5} = \left\{ \btau^x\otimes \Itwo, \btau^z\otimes \Itwo, \btau^y\otimes \bsig^x, \btau^y\otimes \bsig^y, \btau^y\otimes \bsig^z \right\}
\eeq
The ten elements $\Lambda^m$, $m=1,\ldots,10$ can be obtained from the
commutators $[\Gamma^{n},\Gamma^{n'}]/(2i), n=1,\ldots,5, n' > n$. The
grouping of these matrices into $\Gamma$s and $\Lambda$s is motivated
by the fact that under the action of the time reversal operator
$\bTheta=-i(\Itwo\otimes \sg^y)K$ where $K$ is the complex conjugation
operator\cite{GottfriedBook}, $\bTheta^{-1} \Gamma^n \bTheta=
\Gamma^n$ while $\bTheta^{-1} \Lambda^m \bTheta = - \Lambda^m$. From the 
fact that the Hamiltonian in \eqn{eqn:TBkspace} is time reversal invariant, and from
the properties of the $\Gamma$ and $\Lambda$ matrices just mentioned,
we get from \eqn{eqn:Hexpand} that\cite{KanePRL05a}
\beq\mylabel{eqn:TRcondition}
\begin{split}
d_n(-\bk) &= d_n(\bk) \\
e_m(-\bk) &= -e_m(\bk)
\end{split}
\eeq
Eqn.~\ref{eqn:TBkspace} along with eqns.~\ref{eqn:Hexpand} and
\ref{eqn:TRcondition} describes a general four-band Hamiltonian
with time reversal symmetry.

The systems of interest are those which possess a gap in their energy
dispersion -- two bands and separated from the other two by an energy
gap -- and the chemical potential lies in this gap. The nature of
this insulating state (topological or trivial) is determined by the
topological properties of the occupied bands and is characterized by
the $Z_2$ index.\cite{KanePRL05a,KanePRL05b,BernevigSCI06,BalentsPRB2007,
RoyPRB2009a,RoyPRB2009b} While
our formulation is applicable to any four-band system with time reversal
symmetry, we shall focus on topological insulators
which possess protected edge/surface states.

\section{Continuum Theory, Variational Principle and Boundary Conditions}
\mylabel{sec:ContinuumTheory}

The continuum theory is developed by focusing on a region of the
Brillouin zone, i.~e., specifically around the $\bk$-points which
support low energy excitations. In the case of topological insulators
with a bounding edge (or surface), the low energy excitations (i.~e.,
excitations close to the chemical potential) usually occur near a time
reversal invariant momentum (TRIM).\cite{HasanRMP10} TRIMs occur
at the origin of the Brillouin zone, at the zone edges etc. In what
follows, we shall develop the continuum theory focusing on the $\bk=\bzero$
TRIM; generalization to any other TRIM of interest is straightforward.

We discuss the continuum theory in the first quantized form. For our
four-band model, the wave function is a four component
vector function $\psi_a(\bx)$ of the position $\bx$. We look to determine a Hamiltonian operator
$\Hds$ that dictates the time evolution of $\psi_a(\bx)$, i.~e.,
\beq \mylabel{eqn:ContinuumScrodinger}
i \dot{\psi}_a(\bx) = \Hds_{ab} \psi_b(\bx)
\eeq
where the dot represents time derivative and the repeated index $b$ is
summed over. We have set $\hbar=1$. To determine $\Hds$, we expand the function $d_n(\bk)$
and $e_m(\bk)$ about $\bk = \bzero$ up to quadratic order, which upon
using \eqn{eqn:TRcondition} gives
\beq
\begin{split}
d_n(\bk) & = d_n^0 + k_i S^n_{ij} k_i \\
e_m(\bk) & = 2 A^m_i k_i
\end{split}
\eeq
where the constants $d_n^0$, tensors $S^n_{ij}$ and vectors $A^n_i$
are properties of the four-band system that characterize the
dispersion near $\bk = \bzero$. We thus have
\beq
H_{ab}(\bk) \approx  (d_n^0 +  S^n_{ij} k_i k_j) \Gamma^n_{ab} + 2A^m_i k_i \Lambda^m_{ab}
\eeq
where repeated $n$ and $m$ indices are summed over the ranges
indicated in \eqn{eqn:Hexpand}. $\Hds_{ab}$ can now be obtained as
$\Hds_{ab} = H_{ab}(k_i \rightarrow -i \dou_i)$ where $\dou_i \equiv
\dou/\dou x_i$, i.~e.,
\beq
\Hds_{ab} =   (d_n^0 -  S^n_{ij} \dou_i \dou_j) \Gamma^n_{ab} - 2 i A^m_i \dou_i \Lambda^m_{ab}.
\eeq
which completes the discussion of \eqn{eqn:ContinuumScrodinger}. 

Consider now a region of space (in two or three dimensions) $\Omega$
bounded by a boundary $\dou \Omega$ (which may be an edge or a surface). The
stationary states at low energy are eigenstates of the continuum
Hamiltonian $\Hds$, i.~e.,
\beq\mylabel{eqn:stationary}
\Hds_{ab}\Psi_b(\bx) = E \Psi_a(\bx)
\eeq
where $E$ is the energy eigenvalue, with appropriate boundary
conditions for the four component wavefunction $\Psi_a(\bx)$ on $\dou
\Omega$.

To aid the determination of the boundary conditions, here we propose
an energy functional associated with a four component wavefunction
$\Psi_a(\bx)$:
\begin{widetext}
\begin{equation}\label{eqn:EnergyFunc}
  {\cal E}[\Psi^*(\br),\Psi(\br)] = \int_\Omega \text{d}^d\vec{r}
  \Bigl(  \Psi^*_ad_{n}^0\Gamma^{n}_{ab}\Psi_b  
  -(\dou_i\Psi^*_a)S^n_{ij}\Gamma^{n}_{ab}(\dou_j\Psi_b)
  -i\left[\Psi^*_a A^m_i\Lambda^{m}_{ab}\dou_i\Psi_b + 
          (\dou_i\Psi^*_a)A^m_i\Lambda^{m}_{ab}\Psi_b \right] 
     - E\Psi^*_a\Psi_a \Bigr)
\end{equation}
where $E$ is a Lagrange multiplier that ensures that the wavefunction
is normalized to unity. All repeated indices are summed over their
appropriate ranges.  We now show that the states that render this
energy functional extremal are the stationary states of
\eqn{eqn:stationary}. Towards this end, upon varying $\Psi^*$ by
$\delta \Psi^*$, we get
\beq
\begin{split}
\delta {\cal E} & = \int_\Omega \D{^d \br} \Bigl( (\delta \Psi^*_a)  d_{n}^0\Gamma^{n}_{ab}\Psi_b  
  -(\dou_i(\delta \Psi^*_a))S^n_{ij}\Gamma^{n}_{ab}(\dou_j\Psi_b)
  -i\left[ (\delta \Psi^*_a) A^m_i\Lambda^{m}_{ab}\dou_i\Psi_b + 
          (\dou_i (\delta \Psi^*_a) )A^m_i\Lambda^{m}_{ab}\Psi_b \right] 
     - E(\delta \Psi^*_a)\Psi_a \Bigr) \\
& = \int_\Omega \D{^d \br} (\delta \Psi^*_a) 
\Bigl[ 
 \left((d_n^0 -  S^n_{ij} \dou_i \dou_j) \Gamma^n_{ab} - 2 i A^m_i \dou_i \Lambda^m_{ab}  - E \delta_{ab}\right) \Psi_b \Bigr] + \int_{\dou \Omega} \D{^{d-1} \br} (\delta \Psi^*_a) \Bigl[ n_i \left(S^n_{ij} \Gamma^n_{ab} \dou_j \Psi_b + i A^m_i \Lambda^m_{ab} \Psi_b\right) \Bigr]
\end{split}
\eeq
where we have used the divergence theorem and $n_i$ is the outward
normal to the boundary $\dou \Omega$.
\end{widetext}
The extremality of ${\cal E}$ necessitates that 
\beq
 \left((d_n^0 -  S^n_{ij} \dou_i \dou_j) \Gamma^n_{ab} - 2 i A^m_i \dou_i \Lambda^m_{ab}  - E \delta_{ab}\right) \Psi_b = 0
\eeq
in $\Omega$ which is exactly the stationary Schr\"odinger equation of
\eqn{eqn:stationary}. Further on the boundary $\dou \Omega$, we have
either
\beq\mylabel{eqn:Fixed}
\delta \Psi^*_a = 0
\eeq
which corresponds the {\em fixed} boundary condition where the values
of the wavefunction $\Psi_a$ is fixed (usually to zero), or
\beq\mylabel{eqn:Natural}
 n_i \left(S^n_{ij} \Gamma^n_{ab} \dou_j \Psi_b + i A^m_i \Lambda^m_{ab} \Psi_b\right) =0
\eeq
which is the {\em natural} boundary condition (note, again, that all
the repeated indices are summed). We emphasize that this boundary
condition is applicable to any time reversal invariant four-band
system in two or three dimensions. In particular, the formulation is
tailor made for the study of edge (surface) states of topological
insulators. In the next section, we illustrate this framework by
calculating (analytically) the edge states of a topological insulator
described by the well known BHZ model\cite{BernevigSCI06}.
 
\section{BHZ Model: Comparison of Continuum Theory and Tight Binding Results}
\mylabel{sec:BHZ}

The BHZ model\cite{BernevigSCI06} describes 2D topological insulators
realized in the HgTe/CdTe quantum wells.  The tight binding version of
the model is obtained by considering four spin-orbit coupled orbitals-
$\ket{s\up}$, $\ket{p\up} \equiv \ket{\left(p_y+ip_x\right)\up}$,
$\ket{s\dn}$, and $\ket{p\dn} \equiv \ket{\left(p_y-ip_x\right)\dn}$
per site on a square lattice whose lattice spacing $a$ is taken as
unity.  The model can be written as,
\begin{align}
{\cal H} = \sum_{I\alpha\sigma}\eps_\alpha\cidag{I\alpha}\ci{I\alpha} - \sum_{I\bm{\delta}\alpha\beta\sigma}
t_{\alpha\beta}(\bdelta \sigma) \cidag{(I+\bdelta)\alpha}\ci{I\beta}
\label{eq:bhzmodel}
\end{align}
where $\alpha, \beta = s, p$ and $\eps_{\alpha}$ denote the orbital
energies.  $\sigma = \up, \dn$ and $\bm{\delta}$ is a nearest
neighbour vector.  The hopping matrix elements
$t_{\alpha\beta}(\bm{\delta}\sigma)$ in the $\ket{s\sigma}$,
$\ket{p\sigma}$ basis are given by,
\begin{align}
t(\pm\hat{x}\sigma) = \begin{pmatrix}
t_{ss} & \pm\sigma\frac{it_{sp}}{\sqrt{2}}  \\
\pm\sigma\frac{it_{sp}}{\sqrt{2}} & -t_{pp} \\
\end{pmatrix},
\;\;
t(\pm\hat{y}\sigma) = \begin{pmatrix}
t_{ss} & \pm\frac{t_{sp}}{\sqrt{2}}\\
\mp\frac{t_{sp}}{\sqrt{2}} & -t_{pp}\\
\end{pmatrix}
\end{align}
where $t_{ss}$, $t_{sp}$, $t_{pp}$ are overlap integrals and $\sigma =
+1$ ($-1$) for spin $\up$ ($\dn$).  In the reciprocal space, as in
\eqn{eqn:TBkspace}, this Hamiltonian is described by matrices
\begin{align}
   H(\vec{k}) = \begin{pmatrix} h(\vec{k}) & 0 \\ 0 & h^*(-\vec{k}) \end{pmatrix}
\end{align}
where 
\begin{align}
h(\vec{k}) = \begin{pmatrix}
  \eps_{s} - 2t_s\left(\cos k_x+\cos k_y\right) & 2t_{sp}(\sin k_x-i\sin k_y) \\
  2t_{sp}(\sin k_x+i\sin k_y) & \eps_{p} + 2t_s\left(\cos k_x+\cos k_y\right)
\end{pmatrix}
\end{align}
where we have set $t_{ss} = t_{pp} = t_s$. Further defining $\eps_0$
such that $\eps_s= -(\eps_0-4t_s)$ and $\eps_p = (\eps_0-4t_s)$ we
have
\beq\mylabel{eqn:BHKLamdaGamma}
H(\bk) = d_2(\bk) \Gamma^2 + e_1(\bk) \Lambda^1 + e_2(\bk) \Lambda^2
\eeq
in the form of \eqn{eqn:Hexpand}, with $\Gamma^2 = \btau^z \otimes \Itwo,
\Lambda^1 = \btau^x \otimes \bsig^z, \Lambda^2 = \btau^y \otimes
\Itwo$ and
\beq\mylabel{eqn:BHZdsAndes}
\begin{split}
d_2(\bk) &= -\eps_0 + 2t_s \left(2 - (\cos k_x + \cos k_y) \right) \\
e_1(\bk) &= 2 t_{sp} \sin{k_x} \\
e_2(\bk) &= 2 t_{sp} \sin{k_y}
\end{split}
\eeq
All other $d(\vec{k})$-s and $e(\vec{k})$-s are zero. Note that here we have relabelled
the $m$ index in \eqn{eqn:Hexpand} for convenience. With this,
focusing on the TRIM at $\bk=\bzero$, we get the continuum Hamiltonian
operator as
\begin{align}\mylabel{eqn:BHKHds}
  \Hds = \left(-\eps_0 - t_s(\dou_x^2+\dou_y^2)\right)\Gamma^2 - 2it_{sp}\dou_x\Lambda^{1}
  - 2it_{sp}\dou_y\Lambda^{2}
\end{align}
with $d_2^0 = -\eps_0, S^2_{ij} = t_s \delta_{ij}, A^1_x = t_{sp},
A^2_y = t_{sp}$; all other $d$-s, $S$-s, $A$-s are zero.  
This Hamiltonian, upon setting $t_s = 1$ has two scales,
$\eps_0$ and $t_{sp}$. When $\eps_0 > 0$, the system is in the
topological phase; the remainder of the discussion considers only this
case. The quantity $t_{sp}$ is a measure of the hybridization of the
$s$ and $p$ orbitals and determines the ``multi-componentness'' of the
wavefunctions.  It must be noted that this model conserves the spin
quantum number, i.~e., the $\uparrow$ and $\downarrow$ spins decouple
at the one particle level.

In order to study the edge states of this model, we consider a
geometry with $\Omega = (-\infty,\infty)\times(0,L)$, i.~e., and
infinitely long (along $x$-direction) ribbon of width $L$ (terminated
at $y=0$ and $y=L$, i.~e, $\dou \Omega = (y=0) \cup (y=L)$). When $L \rightarrow \infty$, we get a half-space.

Since the spins sectors decouple, we shall consider only the
$\uparrow$-spin sector; the results of the $\downarrow$-spin sector can be
obtained by a time reversal operation. Exploiting the translational
invariance along the $x$-direction, we write $\Psi_\alpha(x,y) = e^{i
  k x} \Psi_\alpha(y)$. For a given momentum $k$, the functions
$\Phi_\alpha(y)$ satisfy \eqn{eqn:stationary} with $\Hds$ given by
\eqn{eqn:BHKHds}:
\begin{align}
  \begin{pmatrix} -\epsilon_0+t_s(k^2-\partial_y^2) & 2t_{sp}(k-\partial_y) \\ 
    2t_{sp}(k+\partial_y) & \epsilon_0-t(k_x^2-\partial_y^2) \end{pmatrix}
\begin{pmatrix} \Psi_s \\ \Psi_p \end{pmatrix} = E\begin{pmatrix} \Psi_s \\ \Psi_p 
\end{pmatrix}
\end{align}

Defining $\Phi = \Psi_s + \Phi_p$, $\Psi = \Psi_s - \Psi_p$,
$G(D)\equiv G(\dou_y)=-\epsilon_0+t_s(k^2-\dou_y^2)$ and $H(D)\equiv
H(\dou_y)=-2t_{sp}\dou_y$, we get
\beq\mylabel{eqn:PhiPsi}
\begin{split}
  G(D)\Psi - H(D)\Psi &= (E-2t_{sp} k)\Phi \\
  G(D)\Phi + H(D)\Phi &= (E+2t_{sp} k)\Psi
\end{split}
\eeq
which leads to 
\begin{align}
  [G(D) - H(D)][G(D) + H(D)] \Phi &= (E^2 - 4t^2_{sp} k^2)\Phi
\end{align}
Assuming a trial solution $\Phi(y)=e^{qy}$, we obtain the 
following quartic equation for $q$, 
\begin{align}
  t_s^2(k^2-q^2)^2 + 2(-\eps_0t_s + 2t_{sp}^2)(k^2-q^2) + (\eps_0^2-E^2) = 0 
\end{align}
which gives four solutions for $q$,  $q_{1,2} = \pm q_I$, $q_{3,4}=\pm q_{II}$ which are given by,
\begin{align}\mylabel{eqn:qsols}
  q_{I,II}^2 = k^2+\frac{(-\eps_0t_s + 2t_{sp}^2)\pm \sqrt{4t_{sp}^2(t_{sp}^2 - \eps_0t_s)
  + t_s^2E^2 }}{t_s^2}
\end{align}
Therefore the general solution for $\Phi$ and $\Psi$ are given by,
\begin{align}
  \Phi(y) &= {\cal A}_1e^{q_1y} + {\cal A}_2e^{q_2y} + {\cal A}_3e^{q_3y} + {\cal A}_4e^{q_4y} 
  \mylabel{eqn:genphi}
  \\
  \Psi(y) &= \frac{1}{E+2t_{sp} k}\left\{G(D)+H(D)\right\}\Phi(y)
  \mylabel{eqn:genpsi}
\end{align}
where ${\cal A}_i$-s are four constants. 
The complete solution for the wavefunction is given by,
\begin{align}
  \Psi_\alpha(x,y) \equiv \begin{pmatrix}\Psi_s\\ \Psi_p\end{pmatrix}e^{ikx}
    = \frac{1}{2}\begin{pmatrix}\Phi+\Psi\\ \Phi-\Psi\end{pmatrix}e^{ikx}
  \label{eqn:gen_wf}
\end{align}

The determination of the energy eigenvalue $E$ and the constants ${\cal A}_i$-s
requires the boundary conditions. The fixed boundary
condition\cite{ZhouPRL08} \eqn{eqn:Fixed} reads
\beq\mylabel{eqn:BHZFixed}
\begin{split}
\Psi_s(0) = \Psi_p(0) &= 0 \\
\Psi_s(L) = \Psi_p(L) &= 0 
\end{split}
\eeq
while the natural boundary condition derived in \eqn{eqn:Natural} provides
\beq\mylabel{eqn:BHZNatural}
\begin{split}
  t_s\frac{\text{d}\Psi_s}{\text{d}y} + t_{sp}\Psi_p &= 0 \\
  t_s\frac{\text{d}\Psi_p}{\text{d}y} + t_{sp}\Psi_s & = 0
\end{split}
\eeq
on $\dou \Omega$ i.e., at $y=0$ and $y=L$.

\begin{figure}
 \centering
  \includegraphics[width=\myfigwidth]{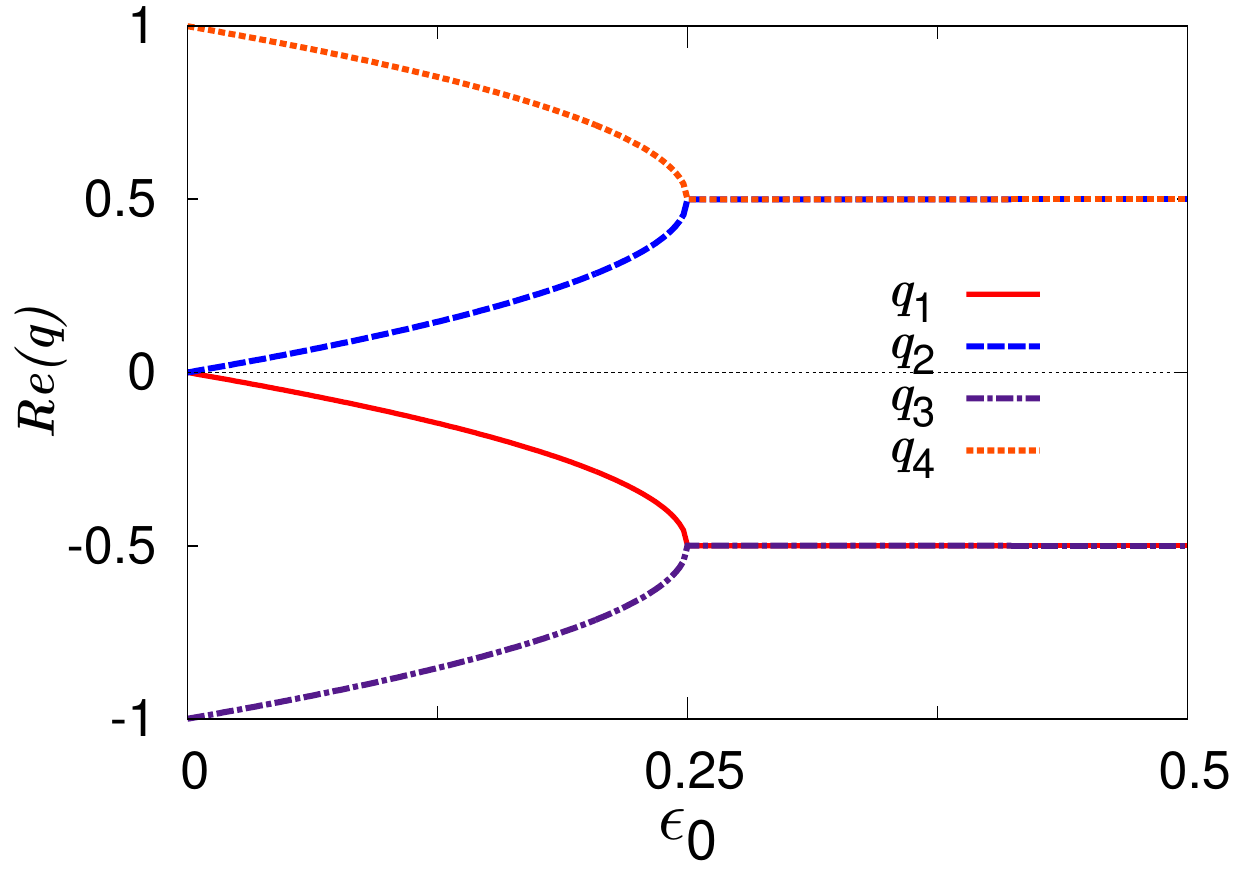}\hfill
  \includegraphics[width=\myfigwidth]{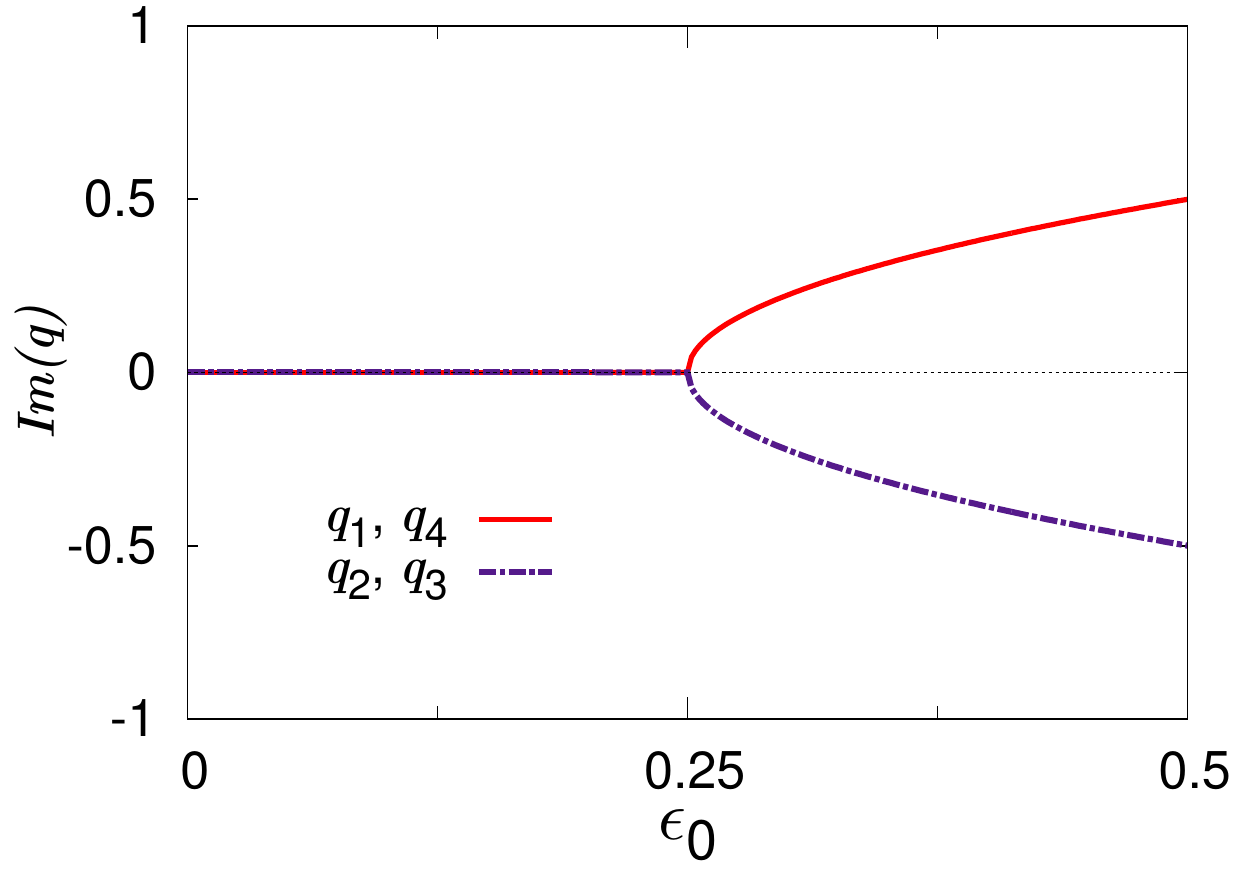}
  \caption{The dependence of the wavevectors $q$ that determine the
    nature of the edge states on $\eps_0$. Top: Real part of
    $q$. Bottom: Imaginary part of $q$. For $\eps_0 < t_{sp}^2$, the
    edge states are exponentially decaying, while for $\eps_0 >
    t_{sp}^2$, they have an oscillating character along with the
    exponential decay. }
  \mylabel{fig:qval}
\end{figure}

\begin{figure}
 \centering
  \includegraphics[width=\myfigwidth]{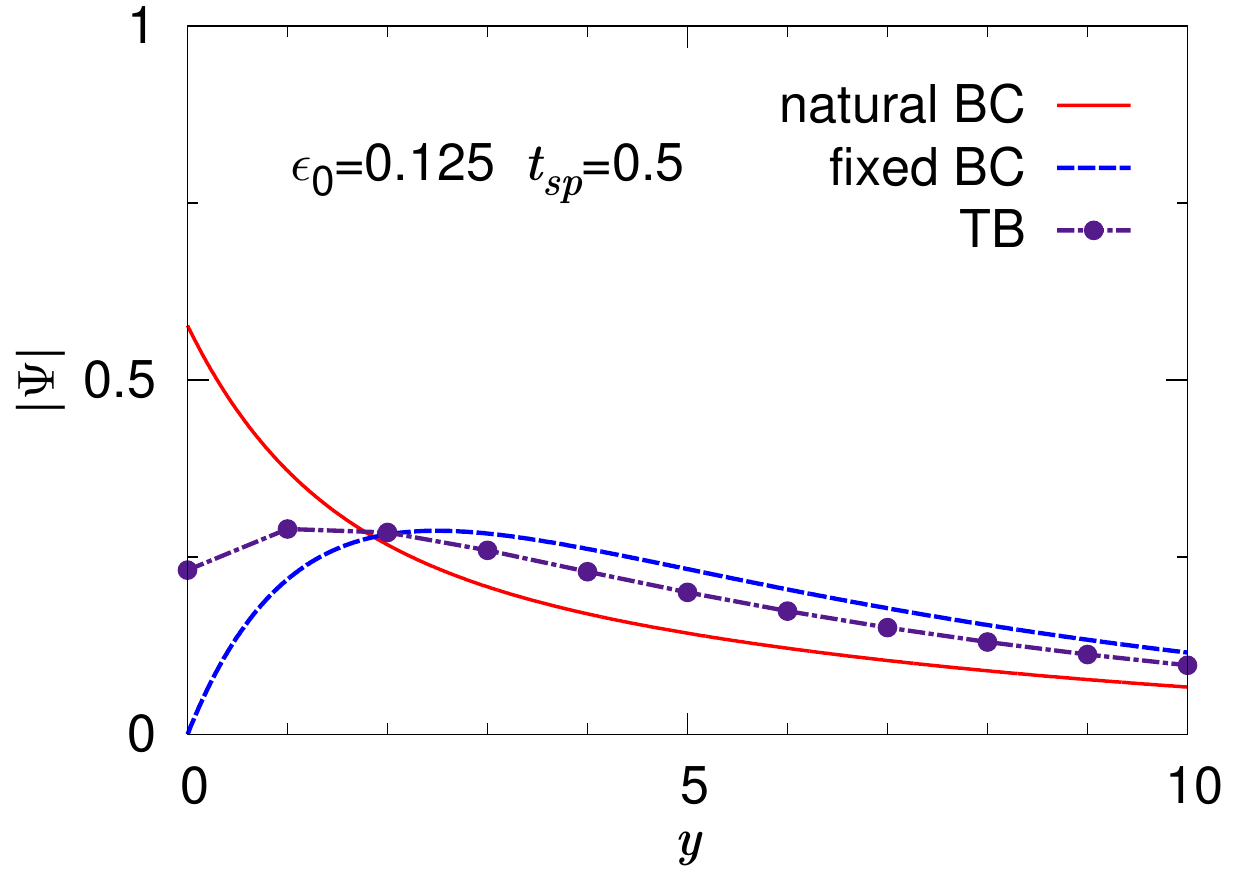}\hfill
  \centerline{(a)} \\
  \includegraphics[width=\myfigwidth]{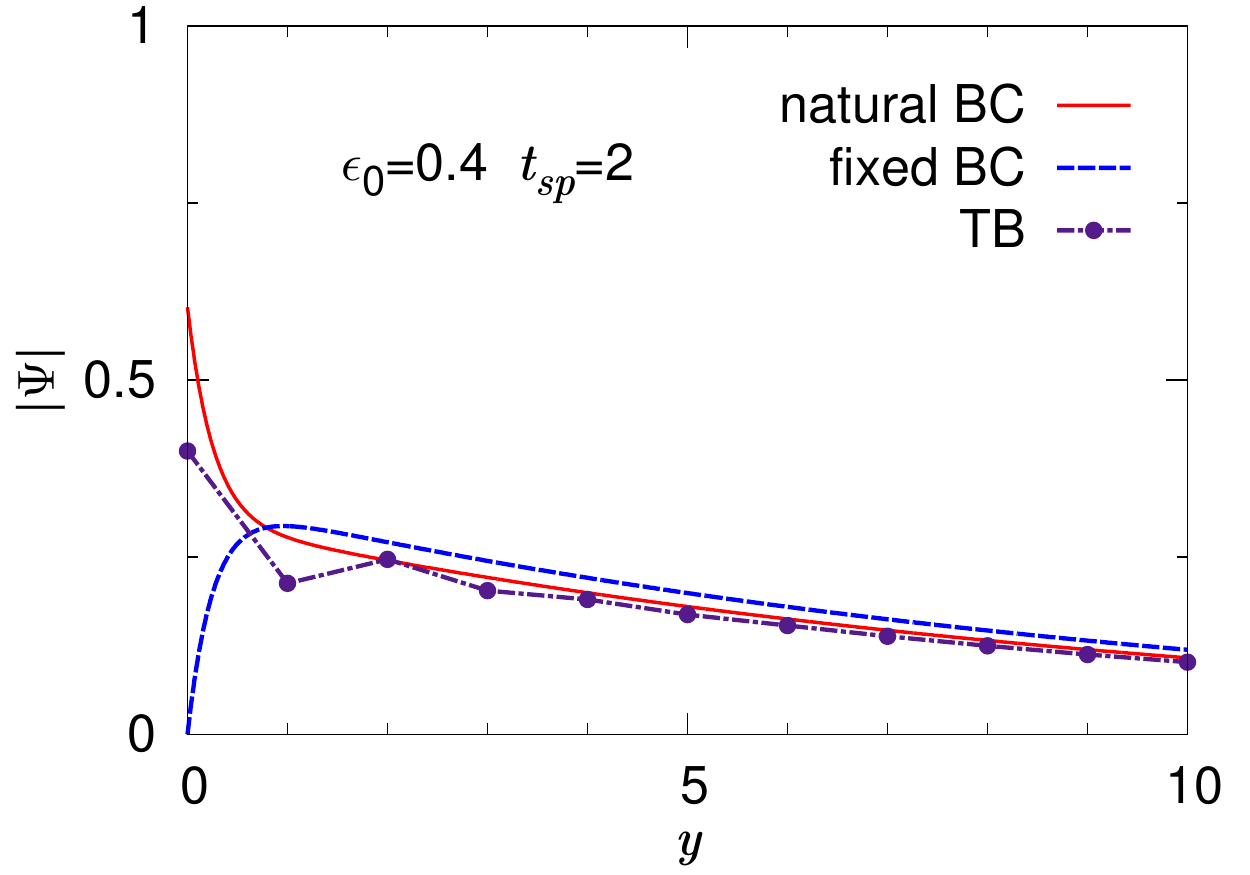}\\
  \centerline{(b)} 
  \includegraphics[width=\myfigwidth]{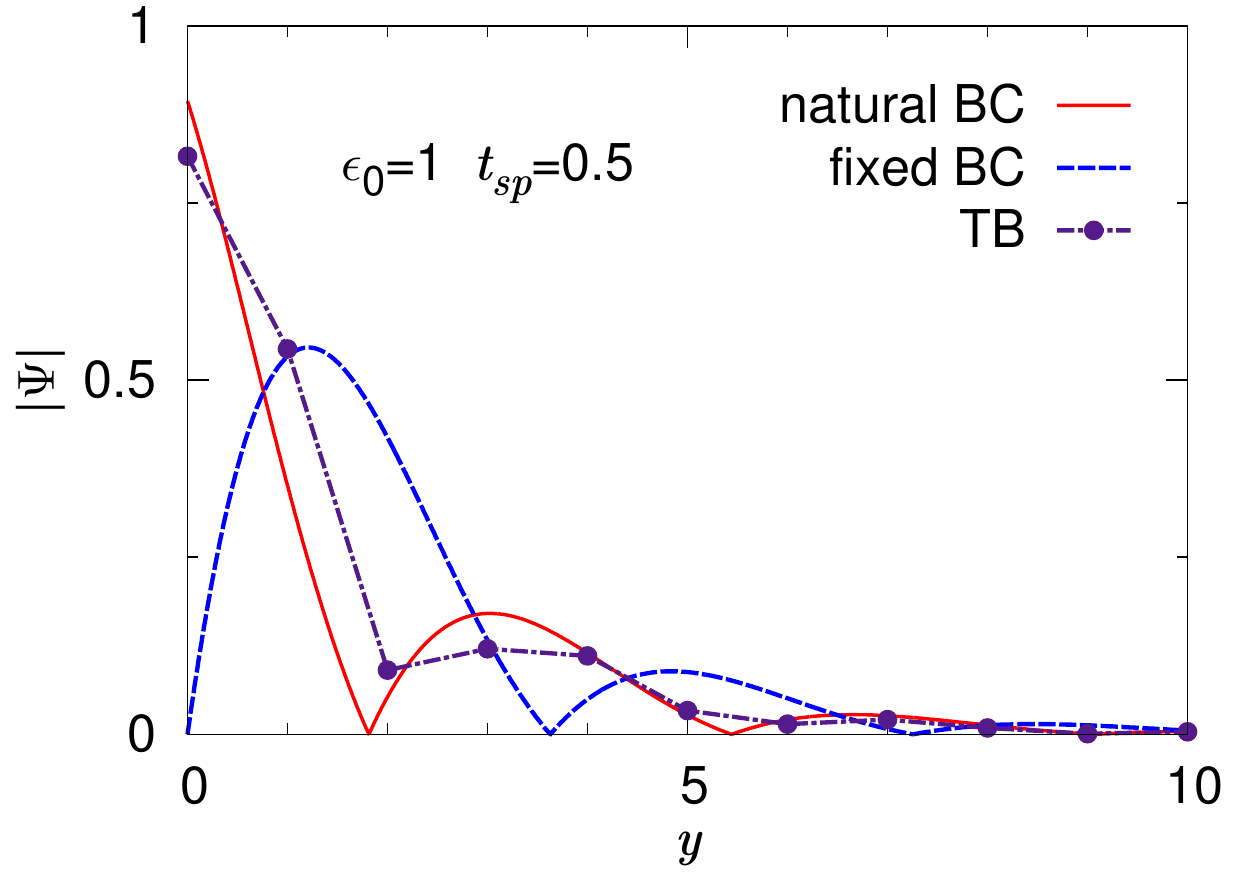}\\
  \centerline{(c)} 
  \caption{Comparison of the edge state wavefunctions obtained
    analytically by using the two different BCs with the corresponding
    result from the tight binding calculation. The wavefunctions plotted are for $k=0$; only the $\Psi_s$ component is shown.}
  \mylabel{fig:compareWf}
\end{figure}

In the remainder of the discussion $t_s$ is set to unity. It is useful
to discuss the nature of the solution of $q$, before proceeding to
compare the analytical results with the numerical tight binding
calculations. Note that the values of $q$ depends on the energy
eigenvalue $E$ (\eqn{eqn:qsols}). Since the $\bk=\bzero$ corresponds
to a TRIM, we expect pair (time reversal related) of topologically
protected edge states at $k=0$, and by the symmetry of the problem, we
expect $E=0$ to be their energy eigenvalue. The values of $q$ with
$E=0$ are then determined by the parameters $\eps_0$ and $t_{sp}$,
i.~e., they are characterized by the same parameters that determine
the ``topology'' of the system.  \Fig{fig:qval} shows a plot of the
$q$s as a function of the parameter $\eps_0$. We find that there are
two regimes of $\eps_0$, $\eps_0 < t_{sp}^2$, where there are four
distinct real roots for $q$s and $\eps_0 > t_{sp}^2$ where $q$s are
complex and appear in conjugate pairs. In the former regime, magnitudes
of $q_1$ and $q_2$ increase with increasing $\eps_0$, while that of
$q_3$ and $q_4$ decrease with increasing $\eps_0$. In the latter
regime, the real parts of $q$ are unaffected, while their imaginary
parts increase in magnitude. Clearly, the nature of the edge states
for $\eps_0 < t_{sp}^2$ is different from that for $\eps_0 >
t_{sp}^2$. In the former case, the edge state wavefunction is
non-oscillating and falls exponentially as the distance from the
edge. In the latter case, the wave function also has an oscillatory
part, and as we shall show later, this leads to quite interesting
physics and possibilities. 

\subsection{Half Space}
\mylabel{sec:SemiInfinite}

Let us first consider a semi-infinite plane with its boundary at $y=0$. Then the bounded solution for $\phi$ is given by,
\begin{align}
  \phi(y)={\cal A}_2e^{-q_Iy}+{\cal A}_4e^{-q_{II}y}
\end{align}
The energy eigenvalues and the wave functions can be determined by imposing either the fixed boundary condition \eqn{eqn:BHZFixed} or the natural boundary condition \eqn{eqn:BHZNatural}. After some simple algebra, it can be shown that for small $k$,
\begin{align}
  E(k) = 2t_{sp}k
\end{align}
a linear dispersion for the edge states, that is, remarkably, {\em
  independent} of which boundary condition is chosen.

This value of $E(k)$ can be now used to determine the constant
coefficients ${\cal A}_{2,4}$ and hence $\Psi_{s,p}$. The profile of the wave
function, of course, depends strongly on the
boundary conditions. \Fig{fig:compareWf} shows a comparison of the results of the
analytical formulation presented above with the two different boundary
conditions and the wave function obtained from numerical calculations
with the full tight binding model. \Fig{fig:compareWf}(a) shows that
for a value of $t_{sp} = 0.5$, the wave function calculated from with
the fixed boundary condition differs significantly from that of the
tight binding results for points close to the edge (near $y=0$). The wavefunction with the
natural boundary condition does not vanish at the boundary and has 
the expected exponential decay into the bulk. At large distances from
the boundary the tight binding result for the wave function falls
between the analytical results of the fixed and natural boundary
conditions. This can be understood by noting that the fixed boundary
condition kills the weight of the edge state near the boundary, and
hence overestimates the weight of the wave function in the bulk. The
effect is precisely the opposite with the natural boundary condition,
where the weight in the bulk is underestimated compared to tight
binding result. We now consider \Fig{fig:compareWf}(b) which shows the
comparison of the edge state wave function with $t_{sp}=2$, but still
with $\eps_0 < t_{sp}^2$. In this case we see that the wavefunction
determined by the natural boundary condition not only closely
reproduces the qualitative aspects of the tight binding solution, but
is also in excellent quantitative agreement with it at large distance
from the edge. Finally, in \Fig{fig:compareWf} we show the comparison
of the wave functions in the regime of parameters with $\eps_0 >
t_{sp}^2$. We see, again, that the analytical wave function obtained
with the natural boundary condition more closely matches the results of tight
binding calculation.

\begin{figure}
\centering
\includegraphics[width=\myfigwidth]{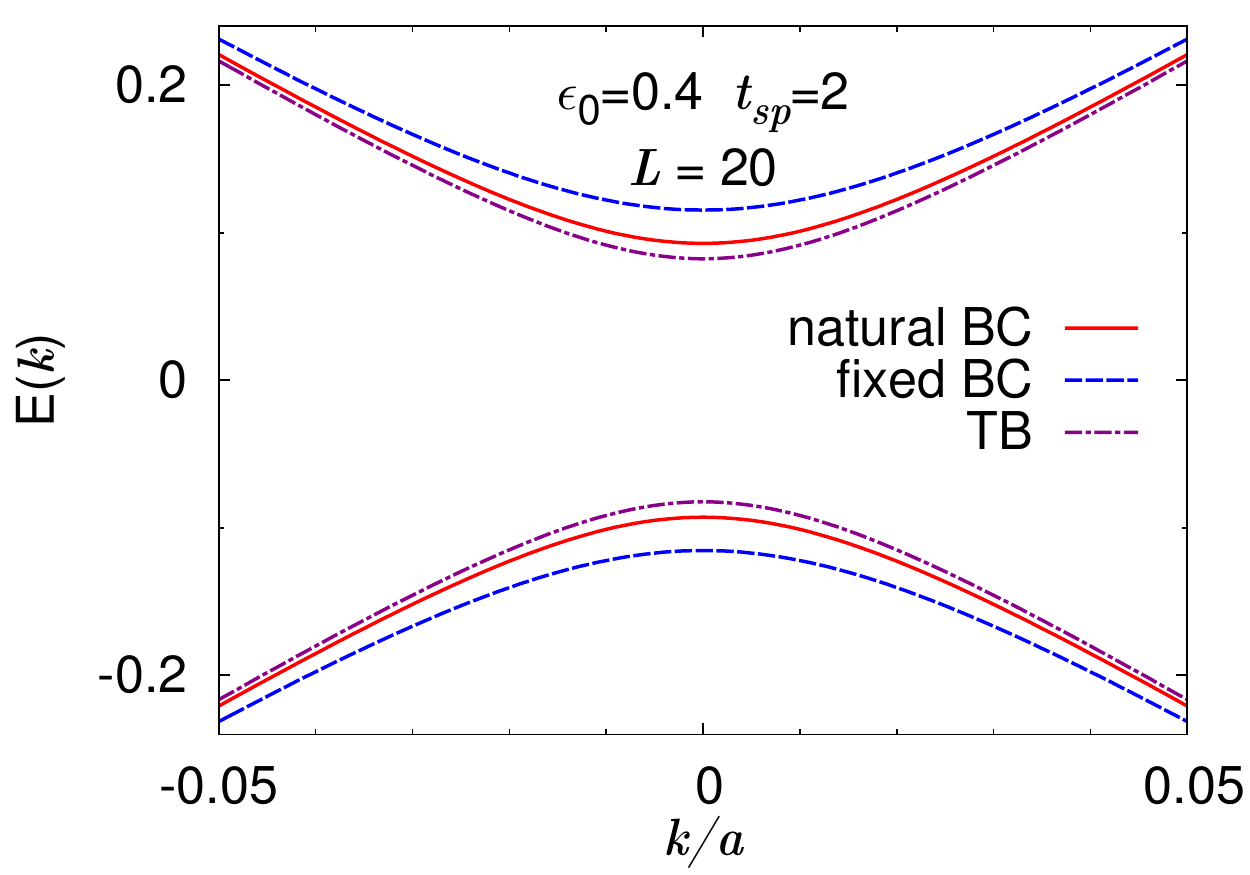}
\caption{Energy dispersion of edge states of a BHZ ribbon of width
  $L=20$. For the parameter values shown, the continuum theory with
  the natural boundary condition (\eqn{eqn:Natural}) reproduces the
  tight binding result more accurately.}
\mylabel{fig:EnergyDispersion}
\end{figure}

\begin{figure}
 \centering
  \includegraphics[width=\myfigwidth]{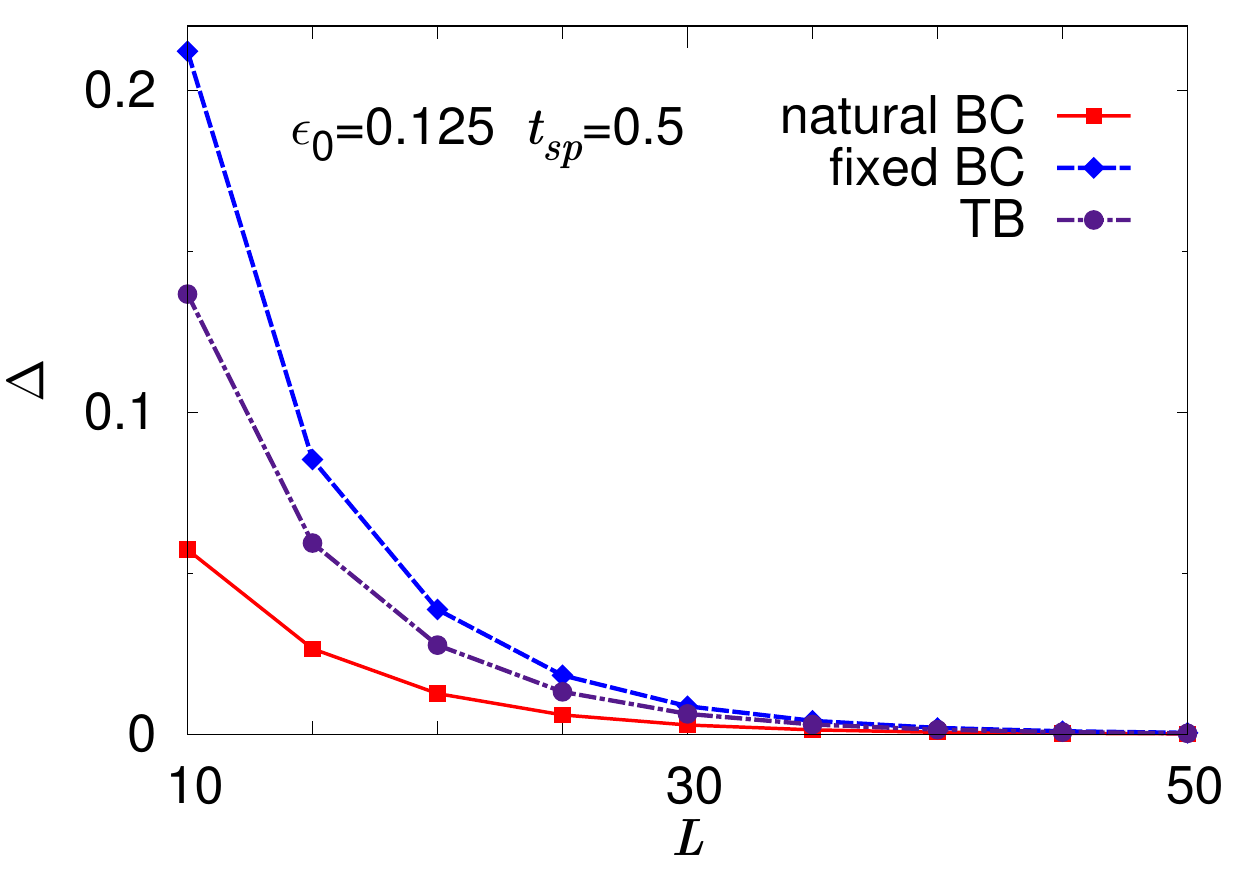}\hfill
  \centerline{(a)}
  \includegraphics[width=\myfigwidth]{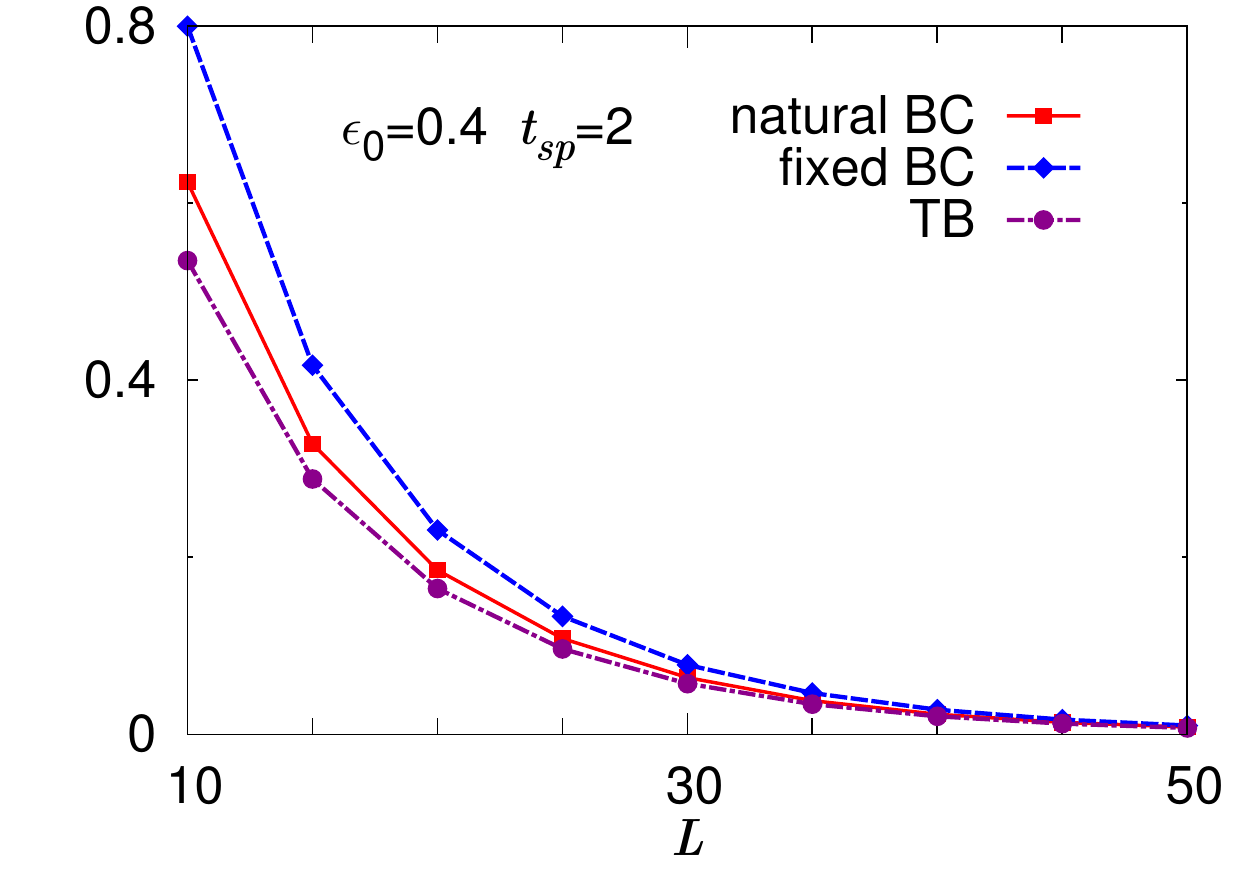}
 \centerline{(b)}
  \caption{Comparison of the energy gap obtained analytically by using the two different BCs 
   with the corresponding result from the tight binding calculation.}
  \mylabel{fig:CompareGap}
\end{figure}

\begin{figure}
 \centering
  \includegraphics[width=\myfigwidth]{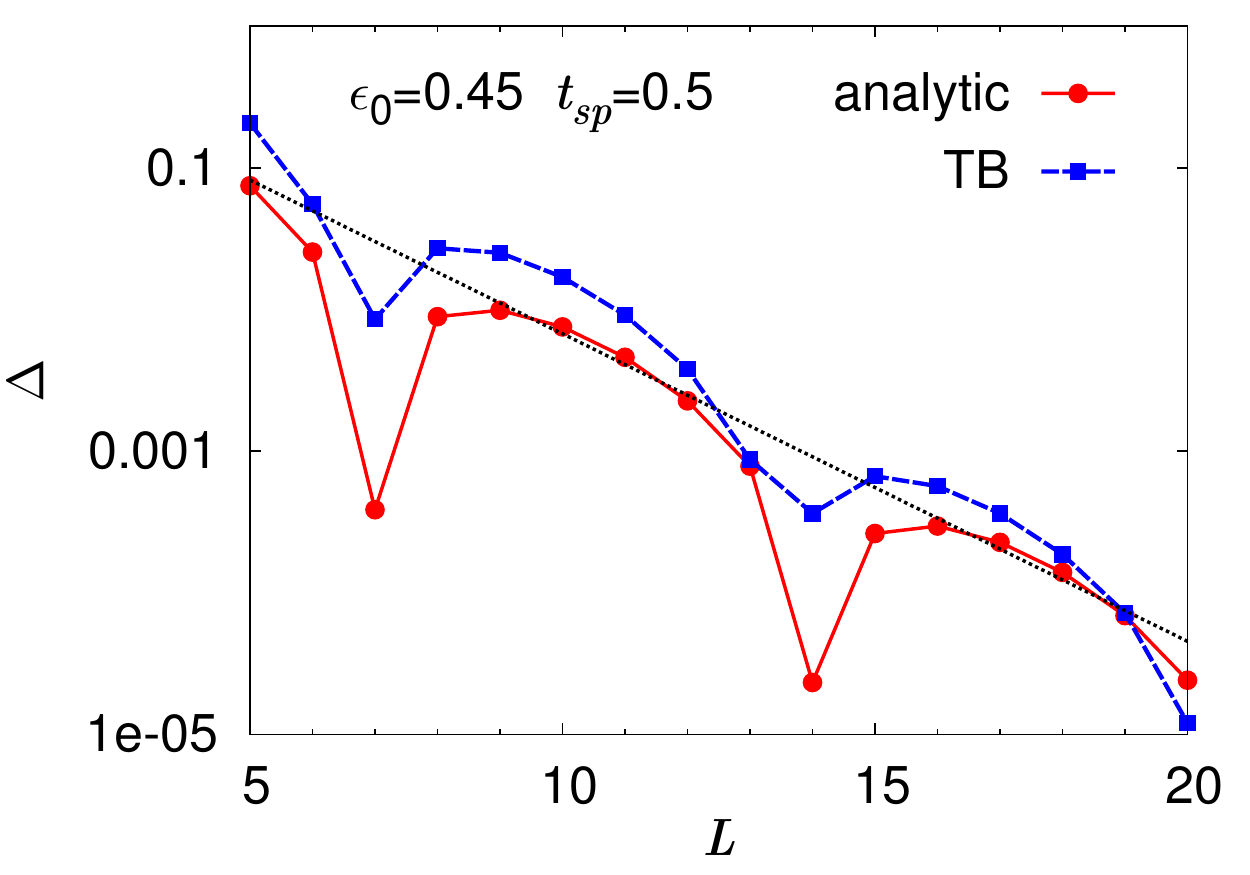}\hfill
  \caption{Non-monotonic dependence of the energy gap with ribbon width $L$ in the parameter regime $\eps_0 > t_{sp}^2$. The dashed line is a guide to the eye to show an overall exponential dependence on the ribbon width.}
  \mylabel{fig:OscGap}
\end{figure}

\subsection{Ribbons}
\mylabel{sec:FiniteRibbons}
We now consider ribbons of finite width $L$. In this case, the edge
states emanating from the edges at $y=0$ and $y=L$, overlap and hybridize
rendering the system gapped (see \Fig{fig:EnergyDispersion}).
A stronger test of the validity of the continuum formulation and the
correctness of the boundary condition can achieved by comparing the
gap calculated using the analytical formulation with that obtained
from the tight binding numerics. \Fig{fig:CompareGap}(a) shows the
comparison of the calculated gaps as a function of the ribbon width
$L$. In this regime of parameters the gap falls exponentially with the
ribbon width as it is determined by the overlap matrix element of the
two edge states emanating from the opposite edges. Again, we see that
in this parameter regime, the tight binding gap lies between the fixed
boundary condition result which is the largest, and the natural
boundary condition value which is the smallest. This can be understood
based on the result of the previous section. The weight of the edge
state wave function in the bulk is overestimated by the use of the
natural boundary condition and hence this gives rise to a larger gap
owing to a larger overlap of the wavefunctions emanating from the
opposite edges. For the same reason, the natural boundary condition
underestimates the gap. For a larger value of $t_{sp}$, the natural
boundary condition is in better quantitative agreement with the tight
binding results. This owes, again to the fact that wave function is
better estimated by the natural boundary condition.

Our final result pertains to the energy gap in ribbons with parameters
in the regime $\eps_0 > t_{sp}^2$. \Fig{fig:OscGap} shows a plot of
the gap as a function of the ribbon width in such a regime; we see
that the gap is {\em non-monotonic}. Although the gap follows an
exponential fall with increasing ribbon width, there are ``magic
widths'' at which the gap is very small; indeed our analytical results
with the natural boundary conditions does reproduce these
features. The physics behind this phenomenon can be traced to the
oscillatory nature of the edge state wave function in this parameter
regime; for some particular widths of the ribbon, there is a ``near
destructive interference'' of the wave functions emanating from the
opposite edges that renders their overlap matrix element small
resulting in a smaller gap. To the best of our knowledge, this is the
first report of such physics in the BHZ model. We believe this is
generic, and in fact, can find possible use in the design nano-scale
devices with topological insulators.

\subsection{Discussion}
As is evident from our results, a continuum field theory with a
natural boundary condition provides an excellent description of systems
with strong ``component-mixing''. In the case of the BHZ model, this will occur
when $t_{sp}$ is large. Physically, in such cases a wave of ``one
flavour'' can be reflected off a boundary as another flavor, and thus
the wave functions do not have to vanish. This applies to the regime
were the wave functions are oscillatory in nature, the current brought
about by one flavour can be reflected in another flavour channel.
They may be contrasted with systems with a single component wave function
such as in a simple ``one component'' tight binding model where the
appropriate continuum boundary condition is that the vanishing of the
wavefunction at the boundary. Topological insulators that are ``deep''
in their topological phase (such as a large $\eps_0$ and $t_{sp}$) are
strongly ``multi-component'' in nature. For such systems the natural
boundary condition is more appropriate.

\section{Summary}
\mylabel{sec:Discussion}

In the paper, we have developed a continuum theory that is applicable
to study four-band time reversal invariant systems. We formulate a
variational energy functional and show that the Schr\"odinger equation
in the bulk is the Euler-Lagrange equation of this functional. This
formulation allows us to obtain the natural boundary condition of the
system. We have compared our analytical results with full tight
binding calculation for the BHZ model for the half-space and finite
ribbons. We show that in the interesting topological regime, the
natural boundary condition derived in this paper is more
appropriate. We believe that our continuum formulation and boundary
conditions will be useful in developing theory of devices and
applications of topological insulators, and continuum theory modeling
of experiments such as tunnelling from surface states.  The
non-monotonic dependence of the gap on the width of a BHZ ribbon is of
particular interest; we believe such features are generic and can have
numerous applications.

\subsection*{Acknowledgement}
AM acknowledges support from  CPDF programme at IISc, Bangalore.  VBS thanks DST (Ramanujan grant) and DAE (SRC grant) for generous support.

\bibliography{bibliography_bc_revisted}

\end{document}